\documentclass[%
 reprint,
 superscriptaddress,
nofootinbib,
 amsmath,amssymb,
 aps,
 prl
]{revtex4-2}

\usepackage{graphicx}%
\usepackage{dcolumn}%
\usepackage{bm}%
\usepackage{hyperref}%

\usepackage{amssymb}
\usepackage{amsmath,bm}
\usepackage{float}
\usepackage[usenames,dvipsnames]{color}
\usepackage[export]{adjustbox}
\usepackage{gensymb}
\usepackage[normalem]{ulem}
\usepackage{latexsym}

\usepackage{soul,xcolor}

\usepackage{multirow}
\graphicspath{{figures/}} %
\usepackage{tikz}
\usepackage[normalem]{ulem}

\newcommand{\cmmnt}[1]{\ignorespaces}

\newcommand{\ket}[1]{\left|  #1 \right\rangle}

\usepackage{graphicx}%
\usepackage{dcolumn}%
\usepackage{bm}%
\usepackage{hyperref}%

\usepackage{amssymb}
\usepackage{amsmath,bm}
\usepackage[percent]{overpic}

 \newcommand{\beginsupplement}{%
        \setcounter{table}{0}
        \renewcommand{\thetable}{S\arabic{table}}%
        \setcounter{figure}{0}
        \renewcommand{\thefigure}{S\arabic{figure}}%
     }

\date{\today}

\begin{document}

\title{
Bose-Einstein condensation by polarization gradient laser cooling}

\author{Wenchao Xu}
\thanks{These authors contributed equally}
\affiliation{Department of Physics and Research Laboratory of Electronics, Massachusetts Institute of Technology, Cambridge, Massachusetts 02139, USA}
\affiliation{Institute for Quantum Electronics, Department of Physics, ETH Z\"{u}rich, Z\"{u}rich 8093, Switzerland}

\author{Tamara \v{S}umarac}
\thanks{These authors contributed equally}
\affiliation{Department of Physics, Harvard University, Cambridge, Massachusetts 02138, USA}
\affiliation{Department of Physics and Research Laboratory of Electronics, Massachusetts Institute of Technology, Cambridge, Massachusetts 02139, USA}

\author{Emily H. Qiu}
\thanks{These authors contributed equally}
 \affiliation{Department of Physics and Research Laboratory of Electronics, Massachusetts Institute of Technology, Cambridge, Massachusetts 02139, USA}

\author{Matthew L. Peters}
 \affiliation{Department of Physics and Research Laboratory of Electronics, Massachusetts Institute of Technology, Cambridge, Massachusetts 02139, USA}
 
\author{Sergio H. Cant\'{u}}
\affiliation{Department of Physics and Research Laboratory of Electronics, Massachusetts Institute of Technology, Cambridge, Massachusetts 02139, USA}

\author{Zeyang Li}
 \affiliation{Department of Physics and Research Laboratory of Electronics, Massachusetts Institute of Technology, Cambridge, Massachusetts 02139, USA}

 \author{Adrian Menssen}
\affiliation{Department of Physics and Research Laboratory of Electronics, Massachusetts Institute of Technology, Cambridge, Massachusetts 02139, USA}

\author{Mikhail D. Lukin}
\affiliation{Department of Physics, Harvard University, Cambridge, Massachusetts 02138, USA}

\author{Simone Colombo}
\affiliation{Department of Physics and Research Laboratory of Electronics, Massachusetts Institute of Technology, Cambridge, Massachusetts 02139, USA}

\author{Vladan Vuleti\'{c}}
\affiliation{Department of Physics and Research Laboratory of Electronics, Massachusetts Institute of Technology, Cambridge, Massachusetts 02139, USA}

\begin{abstract}
Attempts to create quantum degenerate gases without evaporative cooling have been pursued since the early days of laser cooling, with the consensus that polarization gradient cooling (PGC, also known as ``optical molasses'') alone cannot reach condensation. In the present work, we report that simple PGC can generate a small Bose-Einstein condensate (BEC) inside a corrugated micrometer-sized optical dipole trap. The experimental parameters enabling BEC creation were found by machine learning, which increased the atom number by a factor of 5 and decreased the temperature by a factor of 2.5, corresponding to almost two orders of magnitude gain in phase space density. When the trapping light is slightly misaligned through a microscopic objective lens, a BEC of $\sim 250$ $^{87}$Rb atoms is formed inside a local dimple within 40~ms of PGC.
\end{abstract}

\maketitle

Quantum degenerate gases provide an attractive platform for testing fundamental physics \cite{cronin2009optics,bongs2019taking} and simulating various quantum many-body systems \cite{bloch2008many,bloch2012quantum}. In most experiments, highly efficient laser cooling (Doppler cooling followed by polarization gradient cooling (PGC)) takes an atomic gas from room temperature to sub-mK temperatures~\cite{cohentannoudji1989,chu1992}. At this point, the gas is trapped with a typical occupation per quantum state (phase space density, PSD) of $\sim10^{-6}$, limited by detrimental light-induced processes such as photon reabsorption heating and atom loss. Subsequently, to reach the degeneracy, evaporative cooling~\cite{KetterleDruten1996} is applied. The latter is a robust method requiring only favorable atomic ground state collision properties, however the process is slow and necessarily accompanied by a reduction in atom number.

\begin{figure}[htbp]
\centering
\includegraphics[width=0.48\textwidth]{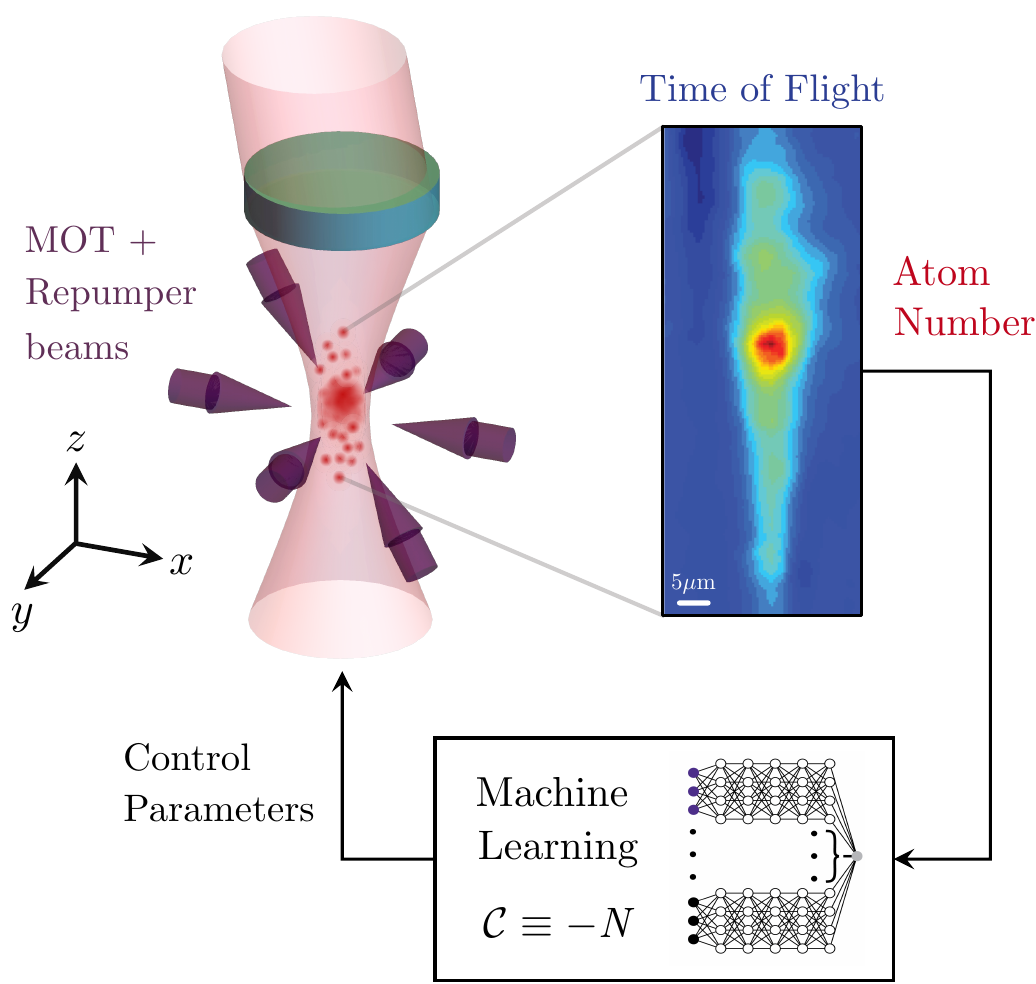}
\caption{\label{fig:Setup}
Schematic of the experimental setup for direct polarization gradient cooling to Bose-Einstein condensation in a corrugated optical potential. The optical dipole trap ($\lambda=808~$nm) is slightly misaligned on a microscope objective lens with numerical aperture $\textrm{NA}=0.4$, resulting in small-scale corrugations of the trapping potential with waist $w_0 \approx 3.5 ~\mu$m.
To maximize the phase space density and reach condensation, machine learning is used to increase the atom number loaded into the trap by tuning the trap loading and cooling parameters (see text). A time-of-flight (TOF) absorption image (right) shows a bimodal momentum distribution. The high-density peak contains 250 out of 2300 atoms.
}
\end{figure}

Recently, alternative optical cooling techniques have been developed that can reach the quantum degenerate regime faster \cite{stellmer2013laser,chen2022continuous,phelps2020subsecond,hu2017creation,urvoy2019direct,Vendeiro2022}. The main obstacle to overcome is light-induced collisional loss at higher atomic densities and within the Bose Einstein condensate (BEC) \cite{Burnett1996}. For Sr atoms, which feature a narrow optical transition, Florian Schreck and colleagues have made use of a strongly inhomogeneous trapping potential to spectrally decouple the emerging BEC from the cooling light \cite{stellmer2013laser}, even demonstrating the first continuous creation of a BEC \cite{chen2022continuous}.
For atoms without a convenient narrow transition, Raman cooling can be employed to mimic a narrower transition using an additional laser field to adjust the effective transition linewidth \cite{chu1992,chu_rsc_2000, salomon1994}. This approach has enabled laser cooling to quantum degeneracy in Rb \cite{hu2017creation,urvoy2019direct} and Cs \cite{solano2019}. However, it has been the general consensus that optical cooling to Bose-Einstein condensation requires relatively sophisticated, finely-tuned techniques, and cannot be accomplished by PGC alone.

In this Letter, we report the direct formation of a Bose-Einstein condensate of $^{87}$Rb atoms using only PGC inside a corrugated potential. The cooling is applied to atoms trapped inside a perturbed optical tweezer, which is formed by intentionally misaligning the trapping light through a high-numerical-aperture microscope objective lens. A small fraction ($\sim 11\%$) of the atoms ($\textit{N} \sim 2500$) forms a BEC, as revealed by a bimodal and anisotropic velocity distribution observed in time-of-flight (TOF) imaging. The BEC formation was first discovered accidentally using a machine learning algorithm designed to maximize the atom number loaded into the misaligned microscopic trap. For traps created by laser beams well-aligned to the microscope objective, we do not observe a condensed component, but we consistently recover condensation in traps misaligned to the microscope. Furthermore, BECs can be generated controllably using a spatial light modulator (SLM) to imprint a speckle-like phase pattern, or by superposing two closely spaced traps. We believe that interference creates dimple structures in the trapping potential that lower the trap depth below the chemical potential of the system, thus facilitating local condensate formation \cite{weber2003bose,kinoshita2005all,xie2018fast,stamper1998reversible}.

Our experimental setup is shown in Fig.~\ref{fig:Setup}. The optical dipole trap is generated with a Gaussian laser beam ($\lambda = 808$~nm) focused to various beam waists ($w_0 = 2-5~\mu$m) through a microscope objective lens (Mitutoyo M Plan Apo NIR B 20X 378--867--5, numerical aperture \textit{NA}~$=0.4$), with the trapping beam slightly tilted ($\sim 3^\circ$) away from normal incidence on the objective.
$^{87}$Rb atoms are loaded into the dipole trap from a magneto-optical trap (MOT). The experimental sequence consists of a 500-ms-long MOT cooling and loading stage, followed by a 40-ms-long MOT compression stage at increased magnetic field gradient, during which atoms are loaded into the dipole trap. The dipole trap depth is typically $U/h = 8$~MHz, with measured radial and axial vibration frequencies of $\omega_r/(2\pi)= 18$~kHz and $\omega_z/(2\pi)= 0.8$~kHz, respectively.

The various parameters (laser intensities and frequencies, bias magnetic fields along $x,y,z$, and magnetic field gradient) for the dipole trap loading during the MOT compression stage were optimized with an open-source, machine learning optimization package (\textsf{M-LOOP})~\cite{Wigley2016mloop}, which has been used previously for improving laser cooling to Bose-Einstein condensation~\cite{Vendeiro2022}. In this work, we use the number of atoms loaded into the dipole trap as our cost function. We divide the 40-ms-long MOT compression stage into three time bins during which we allow \textsf{M-LOOP} to vary 27 parameters in total: the durations of the time bins, the beam powers and detunings of the MOT laser on the $\ket{5S_{1/2}, F=2} \rightarrow \ket{5P_{3/2},F=3}$ transition, and of the repumper laser on the $\ket{5S_{1/2}, F=1} \rightarrow \ket{5P_{3/2},F=2}$ transition, the magnetic fields $B_x, B_y, B_z$, and the gradient of the magnetic quadrupole field. The parameters during the different time bins are connected via linear ramps, except for the laser frequencies, which are jumped in $<1$~ms (see Supplemental Material (SM) \cite{SM} for details). Initial values of all parameters for the MOT compression stage are carried over from the previous MOT loading stage and were chosen by hand.

The optimized sequence obtained by the algorithm (see Fig.~\ref{fig:ControlParameters}) increased the number of loaded atoms by up to a factor of 5 compared to human optimization, while simultaneously decreasing the temperature by a factor of 2.5. To increase the atomic density and reduce light-induced repulsion forces \cite{wieman2000,MOTAnalysisWieman} and losses \cite{Burnett1996}, the algorithm ramps down the intensity of the repumper laser by a factor of $\sim 300$, which transfers most of the atoms into the $F=1$ manifold, and reduces photon scattering. Additionally, the algorithm increases the MOT laser detuning from the $\ket{5S_{1/2}, F=2} \rightarrow  \ket{5P_{3/2}, F=3}$ transition to $-180$~MHz during the final time bin to produce colder temperatures via blue-detuned PGC on the $\ket{5S_{1/2},F=2} \rightarrow \ket{5P_{3/2},F=2}$ transition \cite{salomon2013}. The algorithm further increases the magnetic field gradient to $\sim 25$~G/cm to achieve higher atomic density during dipole trap loading, while the chosen bias magnetic field positions the atomic ensemble at the dipole trap location, maximizing the loading efficiency. All of these are known techniques from human optimization. The algorithm manages to load as many as $N = 2300$ atoms into a microscopic dipole trap with beam waist $w_0 = 3~\mu$m, at a temperature of $T \approx 40~\mu$K. This is remarkable given that under other loading conditions, such traps can be made to load only a single atom~\cite{andersen_review_2016}.

\begin{figure*}[htbp]
\centering
\includegraphics[width=0.99\textwidth]{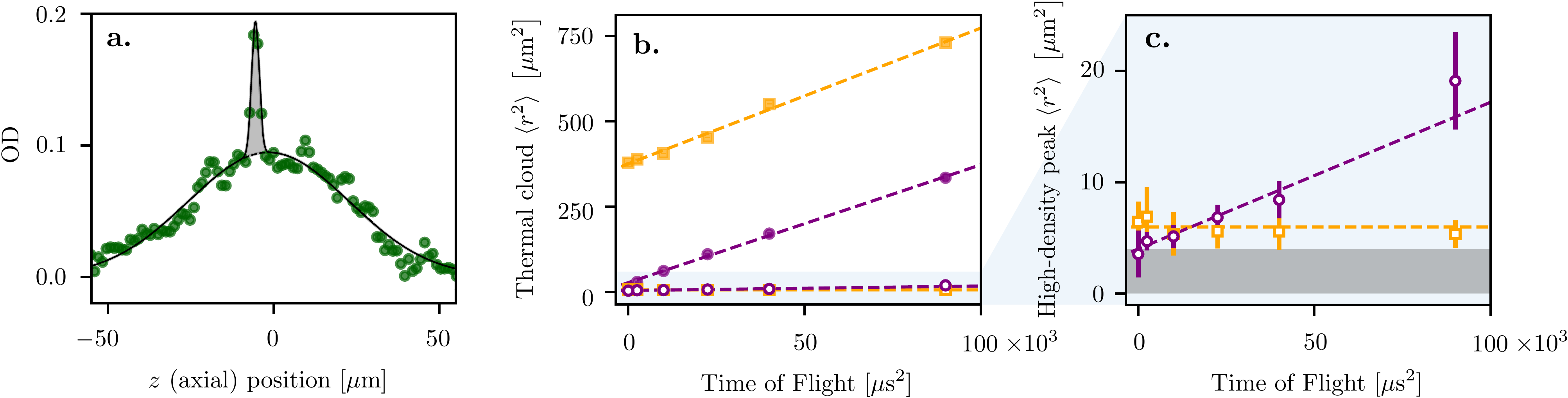}
\caption{\label{fig:Proof}
\textbf{a.} TOF distribution after $400~\mathrm{\mu s}$ shows a high-density peak (gray shaded region) on top of the thermal cloud. \textbf{b.} Variances of the thermal cloud and \textbf{c.} the high-density peak as a function of the TOF. Purple and yellow markers correspond to expansion along the $x$ (transverse) and $z$ (axial) directions, respectively. The thermal cloud fits to temperatures of $T_x=2K_x/k_B=36(2)~\mathrm{\mu K}$ and $T_z=2K_z/k_B=42(2)~\mathrm{\mu K}$ along $x$ and $z$, respectively, with $N=2.7(1)\times 10^3$.
The high-density peak expands anisotropically with much lower kinetic energies of expansion $ 2K^{(0)}_x/k_B = 1.4(2)~\mu$K and $ 2K^{(0)}_z/k_B < 0.2 ~\mu$K. The gray shaded region (below $4~\mu$m$^2$) indicates the imaging resolution.}
\end{figure*}

Following the \textsf{M-LOOP}-optimized trap loading, the atoms are held in the dipole trap for an additional $50 - 300$~ms to allow the ensemble to thermalize, with typical estimated two-body collision rates of $\Gamma_{c} \approx 2.2\times 10^3$~s$^{-1}$ at peak densities $n_0 \approx 5\times 10^{13}~$cm$^{-3}$ of the thermal cloud. Absorption imaging \textit{in-situ} shows a local density peak inside the dipole trap, which indicates a corrugation in the optical potential of the deformed trapping beam. The TOF measurement of the atomic-cloud expansion (Fig.~\ref{fig:Proof}a) reveals that the high-density peak expands anisotropically, and at significantly lower velocities than the thermal cloud. As shown in Fig.~\ref{fig:Proof}b, the thermal cloud expands isotropically in the radial and axial directions with temperatures of $T_x = 2K_x/k_B = 36(2)~\mu$K and $T_z =2K_y/k_B = 42(2)~\mu$K, while the dense peak expands anisotropically with much lower kinetic energies $ 2K^{(0)}_x/k_B = 1.4(2)~\mu$K and $ 2K^{(0)}_z/k_B < 0.2 ~\mu$K. In particular, the expansion in the axial ($z$) direction is too small to be resolved by the imaging system. This anisotropic expansion --- well-below the expansion rate of the thermal gas --- is a hallmark signature of BEC formation \cite{CornellBEC}. A bimodal fit reveals a typical condensate fraction of up to 11\%.

\begin{figure*}[htbp]
\centering
\includegraphics[width=0.95\textwidth]{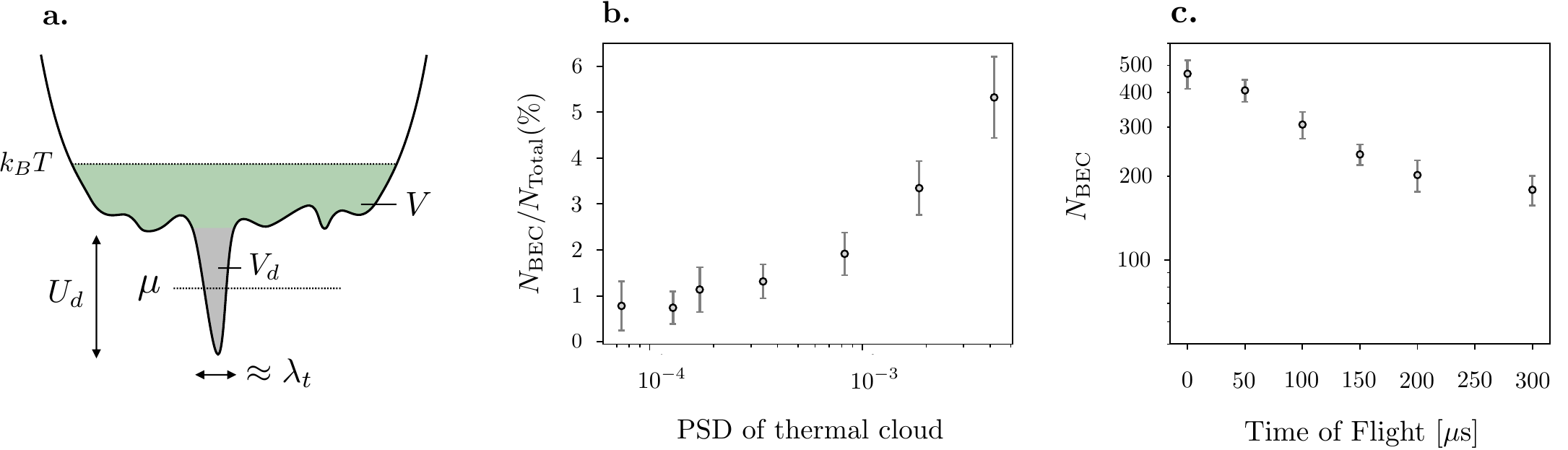}
\caption{\label{fig:ratio_volumes}
\textbf{a.} Corrugations in the optical potential enable BEC formation when the additional trap depth $U_d$ drops below the chemical potential $\mu$. Interference of the trapping beam generates diffraction-limited structures of size $\approx \lambda$, which can exponentially enhance the local PSD$_d$ in the dimple (gray shaded region) with respect to the classical PSD of the thermal cloud (blue shaded region).
\textbf{b.} Fraction of atoms in the BEC peak as a function of PSD in the thermal cloud. The PSD is tuned by ramping up or down the trap power following \textsf{M-LOOP}-optimized dipole trap loading. The data was taken after a TOF of $100\,\mathrm{\mu s}$ for $N=3700(200)$, resulting in up to $200$ atoms in the BEC. \textbf{c.} Atom loss in the condensate inferred from TOF imaging. In-trap, however, the condensate persists for hundreds of milliseconds.
}
\end{figure*}

We hypothesize that interference in the aberrated trapping beam creates corrugations in the potential (local trap dimples), where the local trap potential drops below the chemical potential (see Fig.~\ref{fig:ratio_volumes}a) such that a local condensate can form \cite{weber2003bose,kinoshita2005all,xie2018fast,stamper1998reversible, jacob2011}. 
The stronger confinement of the local dimple relative to the macroscopic trap enhances the local phase space density of atoms inside the dimple. When the volume of the dimple is significantly smaller than that of the macroscopic trap, the phase space density PSD$_{d}$ within the dimple increases exponentially with the depth of the dimple \cite{stamper1998reversible}: 
\begin{equation}
\ln(\textrm{PSD}_d/\textrm{PSD}) =\frac{U_d/{(k_{B}T)}} {1+(V_d/V)e^{U_d/(k_{B}T)}} .  
\end{equation}
Here, $U_d$ is the additional trap depth provided by the dimple, and $V$, $V_d$ are the volumes of the macroscopic and dimple traps, respectively. The expression is an approximation assuming box-shaped trapping potentials and $V_d/V \ll 1$. For smaller volume ratios $V_d/V$, a weaker dimple potential is sufficient to create a BEC. 
The largest possible increase in phase space density is $\textrm{PSD}_d/\textrm{PSD} \sim V/V_d$, obtained for a dimple depth of $U_d/(k_B T) \approx \ln(V/V_d)$. For our parameters, this implies that a volume ratio $V_d/V \lesssim 10^{-3}$ is needed, in combination with an additional dimple potential depth $U_d \approx 6.9 k_B T \approx h\times 5.5$ MHz.
The smallest characteristic structure size of the interference speckle pattern can be estimated as $\lambda/(2 \times \textit{NA}) \approx 1.2 \lambda$, yielding a maximum volume ratio (and hence local PSD increase) of $V/V_d \sim 2 \pi^2 w_0^4/\lambda^4 \approx 1.5 \times 10^3$, matching what is required to explain the BEC formation. For the thermal gas in $V$, the chemical potential calculated from its temperature and atom number is approximately $U_d \approx h \times 5.5$ MHz, satisfying the BEC formation condition.

Additionally, we investigated the dependence of BEC formation on the dipole trapping beam waist $w_0$ and on the misalignment angle between the incident beam and the optical axis of the microscope objective. We found that the trap size is critical: while we observe BEC formation in a dipole trap with $w_0=3.5~\mu$m, no condensation was observed for waists $w_1=2.4~\mu$m or $w_2=4.5~\mu$m. For the smaller beam waist $w_1$, higher light-induced loss during the loading stage \cite{fuhrmanek2012light} into the tighter trap resulted in a smaller loaded atom number $N=800$, and smaller phase space density PSD for the thermal cloud, which was too low to allow BEC formation. On the other hand, for beam waists that are too large, the dimple depth is likely too small to reach the chemical potential (Fig. \ref{fig:ratio_volumes}a).

Once a proper dipole trap beam waist is chosen, the formation of BEC is reproducible, and insensitive to a specific alignment configuration. We verified that BEC formation persists at various angles of misalignment, but the density peak can occur at different locations inside the trap; often, multiple high-density peaks exhibiting anisotropic expansion in TOF are observed (see SM \cite{SM}). In contrast, when the beam is aligned at normal incidence to the microscope objective, we observe only a thermal cloud, without density anomalies or bimodal TOF distributions. Prior to the implementation of the \textsf{M-LOOP} algorithm, BEC formation was not observed, due to smaller loaded atom number and higher temperature obtained from human optimization.

To further verify the BEC formation, we vary the classical phase space density (PSD) of the thermal cloud and measure the corresponding condensate fraction (see Fig.~\ref{fig:ratio_volumes}b). To change the phase space density in the trap, we use the same \textsf{M-LOOP}-optimized loading sequence and subsequently ramp up or down the dipole trap depth within 20~ms, and finally hold the atoms in the trap for an additional 20~ms before TOF imaging. It is evident that higher (lower) PSD yields larger (smaller) condensate fractions, as expected.

In order to test our dimple hypothesis and also create BECs under more controlled conditions, we introduced an SLM in the optical path of the trapping beam.
We were then able to observe on-demand BEC formation for trapping beams well-aligned to the microscope optical axis in two different configurations of the SLM. First, we used the SLM to generate two traps of $2~\mu$m waist and observed BEC formation when the trap separation was in the range $~3.3-3.7~\mu$m, creating interference between the traps. Second, we also observed BEC formation when we introduced a speckle-like pattern consisting of many diffraction orders generated by sine phase functions displayed on the SLM, superimposing three waves with small wavenumbers (and thus small trap spacing in the atom plane) and random phase offsets.
In both cases, we observed persistent BEC formation, but the condensate fractions were smaller than in the data presented in Figs.~\ref{fig:Setup} and~\ref{fig:Proof}. A gallery of different BECs observed in various traps is shown in the SM \cite{SM}.

Notably, while the BEC persists in-trap over timescales of a few hundred milliseconds, we observe atom loss from the condensate during TOF measurements (see Fig.~\ref{fig:ratio_volumes}c). We hypothesize that this atom loss during TOF can be explained by two-body elastic collisions between non-condensed atoms from the larger trap and condensate atoms. In-trap, the atom number in the BEC remains constant because there is a dynamic equilibrium between scattering into and out of the condensate. During TOF, however, the thermal cloud disperses quickly, and loss of atoms from the condensate cannot be recovered via elastic collisions with the thermal atoms. At short TOF, the loss of the condensate may be further enhanced due to the presence of thermal atoms confined within the dimple potential. As the thermal cloud density decreases during TOF expansion, the loss from the condensate stops. A second possible explanation for condensate loss in TOF is three-body loss~\cite{cornell_threebodydecay} (see SM \cite{SM}; note that two-body inelastic collisions should be negligible since the algorithm prepares the atoms in the $F=1$ ground state.) However, the long BEC lifetime in-trap suggests that three-body loss is unlikely to be the dominant loss mechanism on the time scale shown in Fig.~\ref{fig:ratio_volumes}c.

In summary, we have for the first time observed the direct formation of a BEC using only regular optical molasses (PGC) laser cooling, overturning a long-held paradigm that PGC is insufficient for reaching quantum degeneracy. The cooling is accomplished within a duration as short as 40~ms.
Combining this method with arrays of optical tweezers~\cite{endres2016atom,barredo2016atom}, it should be possible to create hundreds of small condensates simultaneously, and purify them by lowering the trap power. This may constitute, e.g., a promising starting point for atom interferometry~\cite{kasevich_chu_1991,mueller_2008} and other precision experiments with ultracold atoms.

We would like to thank Martin Zwierlein and Andrea Muni for insightful discussions, and Zachary Vendeiro for support with \textsf{M-LOOP}. We further thank Shai Tsesses for critical reading of the manuscript. E.H.Q. acknowledges the support of the Natural Sciences and Engineering Research Council of Canada (NSERC). This material is based upon work supported by the U.S. Department of Energy, Office of Science, National Quantum Information Science Research Centers, Quantum Systems Accelerator. Additional support is acknowledged from the NSF-funded Center for Ultracold Atoms, the DARPA ONISQ program,
and ARO.

\bibliography{reference}

\clearpage

\beginsupplement

\section{Supplemental Material}

\subsection{Gallery of observed BECs}

\begin{figure}[htbp]
\centering
\includegraphics[width=0.49\textwidth]{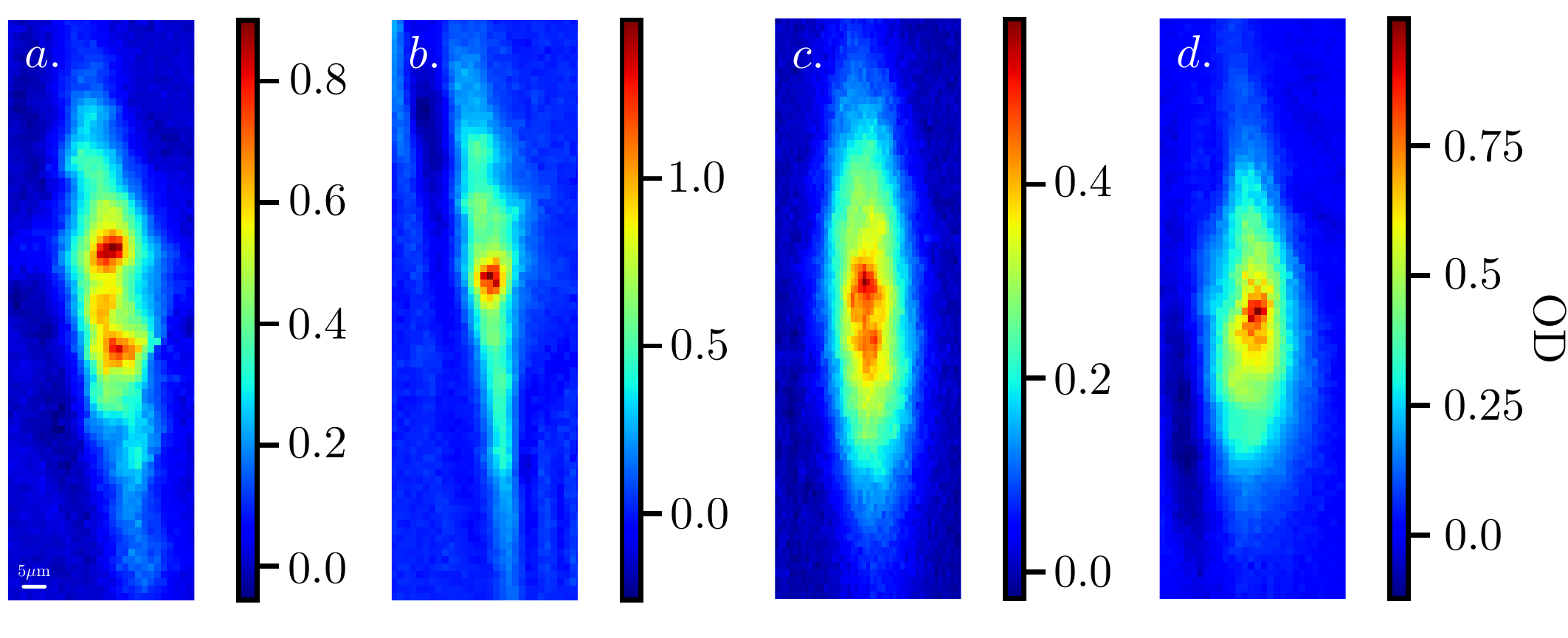}%
\caption{\label{fig:BECGallery}
\textbf{Gallery of BECs.} BEC formation is observed in: \textbf{a.} a misaligned dipole trap with Gaussian beam waist $w_0=3.5~\mu$m (no SLM) after $50~\mu$s TOF. \textbf{b.} a misaligned dipole trap shown in-situ with the trapping beam slightly clipped on the SLM aperture\footnote{The setup consists of a Gaussian beam with $1/e^2$ diameter $2\omega_0 = 12.5~$mm incident on an SLM with aperture size $12.8$~mm$~\times~15.9~$mm. Without accounting for effects of clipping the beam, the Gaussian trap waist is $\omega_0 = 2.5~\mu$m. SLM part number: Hamamatsu x13138-02.
}. This absorption image shows a cloud with $N_{\textrm{BEC}}/N_{\textrm{Total}} = 11\%$. \textbf{c.} a well-aligned dipole trap with a speckle-like phase pattern imprinted on the trapping light using an SLM (see main text for details) after $100~\mu$s TOF. \textbf{d.} a well-aligned trapping potential generated by interfering two traps with waist $\omega_0 = 2~\mu$m separated by $3.5~\mu$m after $20~\mu$s TOF\footnote{The two traps are generated by the SLM in which the difference between blazing angles of the phase pattern maps onto the desired separation on the atom plane.}.}
\end{figure}

Formation of BEC is observed in various trap configurations, as shown in Fig.~\ref{fig:BECGallery}.

\subsection{Machine Learning Optimization}

To maximize the number of atoms loaded into the dipole trap, we use the neural network learner included in the \textsf{M-LOOP} package, which works well for a large number of variables in parameter spaces with many local minima. The neural network learner is comprised of three randomly initialized neural networks, each with 5 hidden layers of 64 neurons. These neural networks, in conjunction with a differential evolution algorithm, learn to map the input parameters to the output cost as summarized in previous works~\cite{Wigley2016mloop, Tranter2018, Vendeiro2022}. 
\begin{figure}[htbp]
\centering
\includegraphics[width=0.48\textwidth,trim= 10 12 7 
8,clip]{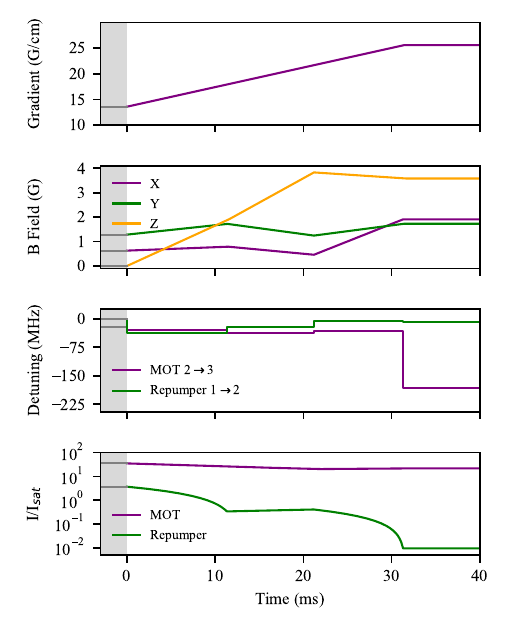}
\caption{\label{fig:ControlParameters}
Optimized control parameters determined by \textsf{M-LOOP}. The initial values shown in the gray shaded region were set by hand (carried over from the prior MOT loading sequence), whereas the colored lines show the values chosen by \textsf{M-LOOP} during the MOT compression sequence. The 27 parameters were allowed to vary independently to minimize the cost function $\mathcal{C} \equiv - \textit{N}$, where $\textit{N}$ is the fitted atom number. Parameters were ramped linearly over three fixed time steps (excluding the detunings). The intensities shown are the sums over all MOT (repumper) beams, where we use $I_{sat}=1.67$ mW/cm$^2$ (Clebsch-gordan for stretch transition).
}
\end{figure}
We choose 27 input parameters for our optimization sequence, consisting of MOT and repumper laser intensities and frequencies, bias magnetic fields along $x,y,z$, and magnetic field gradient. The laser intensities and bias magnetic fields are split into three time intervals, and they are linearly ramped between the chosen values as determined by the algorithm (15 parameters). Due to technical constraints on the laser detunings, the frequencies are jumped discretely without ramping; therefore, we include an extra parameter for each laser detuning at the beginning of the sequence (8 parameters). The detunings are defined with respect to the $\ket{5S_{1/2}, F=2} \rightarrow \ket{5P_{3/2},F=3}$ transition for the MOT laser, and the $\ket{5S_{1/2}, F=1} \rightarrow \ket{5P_{3/2},F=2}$ transition for the repumper laser. We include only a single parameter across the sequence for the quadrupole gradient coil (since the relaxation time of Eddy currents is expected to be comparable to the sequence length), which decreases the complexity of the input parameter space (1 parameter). Finally, we allow the length of each time bin to vary (3 parameters) such that the total duration of the sequence is fixed ($40$~ms). See Fig.~\ref{fig:ControlParameters} for the optimized control parameters.

The \textsf{M-LOOP} package allows the outputs of the learner to be constrained for sparse parameter spaces (or spaces where the signal-to-noise ratio is too low to reliably compute the cost), through the use of a trust region. This feature prevents any parameter from changing by more than a specified fractional amount from the current best value during each new prediction. We choose a trust region of 0.2. Implementation of this feature played a critical role in fast convergence of the algorithm to the optimal parameter, due to the narrow range of parameters that ensures a reasonable number of atoms remain in the trap at the end of the sequence. 

\begin{figure}[htbp]
\centering
\includegraphics[width=0.48\textwidth,trim= 5 0 0 
3,clip]{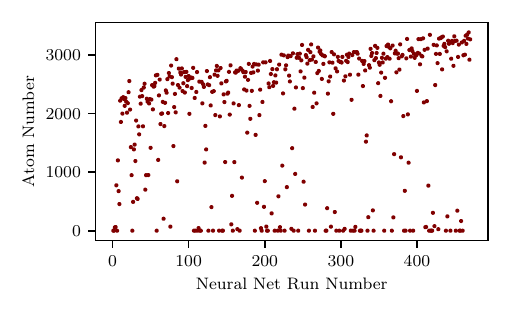}
\caption{\label{fig:Training}
The atom number increases with run number of the neural network learner, implying that the algorithm is learning the parameters that maximize the atom number (or minimize the cost). The atom number is set to zero when the fitted parameters of the atom cloud are unrealistic.}
\end{figure}

For our cost function $\mathcal{C}$, we choose to maximize the number of atoms loaded in the dipole trap, which is obtained from absorption images of the optical depth. Specifically, we choose $\mathcal{C} \equiv - \textit{N}$, where $\textit{N}$ is the fitted atom number. We assume a single run of the experiment successfully loads atoms if the fit parameters of the atom cloud are realistic (i.e. the vertical RMS size is within $15-40~\mu$m, the horizontal RMS size is $<8~\mu$m); otherwise, we assume the fit has failed and we set the cost function to zero to penalize the run. The atom number as a function of run number is shown in Fig.~\ref{fig:Training} for a typical optimization routine.

\subsection{Definition of classical phase space density}

The classical phase space density of the thermal cloud is given by:
\begin{equation}
    \textrm{PSD} = \frac{1}{3} ~N_{\textrm{total}} \frac{\hbar^3 \omega_{r}^2 \omega_{z}}{\left( k_B T \right)^3}
\end{equation}

where $N_{\textrm{total}}$ is the total number of atoms in the dipole trap, $T$ is the fitted temperature of the thermal cloud obtained from TOF imaging, and $\omega_r$ and $\omega_z$ are the radial and axial vibration frequencies, respectively. The factor of $1/3$ accounts for the three Zeeman sublevels of the $F = 1$ ground state manifold.\\

\subsection{Hyperfine states of atoms}
The optimized algorithm prepares the atoms predominantly in the $F=1$ hyperfine manifold by its choice of laser powers and detunings. 
During the final time-bin, the ratio of scattering rates from the MOT laser with respect to the repumper laser (and hence the population ratio in $F=1$ with respect to $F=2$ of the ground state manifold) is calculated as:

\begin{align}
    \frac{P(F=1)}{P(F=2)} &= \frac{I_{\textrm{MOT}}}{I_{\textrm{Repump}}} \left(\frac{\Delta_{\textrm{Repump}}}{\Delta_{\textrm{MOT}}}\right)^2 \nonumber
    \\ &= \frac{22}{9.7 \times 10^{-3}} \left(\frac{-9 \textrm{~MHz}}{84 \textrm{~MHz}}\right)^2 \nonumber
    \\ & = 27,
\end{align}
which confirms that $\sim 97$\% of the atoms are in the $F = 1$ hyperfine manifold of the ground state at the end of the optimized cooling sequence.

\subsection{Dipole trap extinction measurements}
To ensure that the BEC is fully released during TOF, we checked the extinction of the dipole trap power when monitored on a photodiode. In sequence, the extinction is estimated to be $ > 1.7 \times 10^{3}$, limited by the resolution of the oscilloscope used to measure the photodiode voltage. For a typical trap depth of $\sim 8~$MHz, this suggests a residual trap depth of $< 4.7~$kHz during TOF, or $ < 0.2~\mu$K, which is insufficient to confine the BEC along the radial direction. 

\vspace{10pt}

\subsection{Three-body loss in TOF}

The observed condensate expansion energies and atom number imply that the condensate is in the Thomas-Fermi (TF) regime, where the release energy per particle $\varepsilon_r$ is dominated by the interaction energy, such that $\varepsilon_r = 2\mu/7$, where $\mu$ is the chemical potential~\cite{rev_dalfovo}.
Our measurements yield $\mu/h=98(14)$~kHz. From this we can obtain the peak density in the TF regime, $n_{TF}=\mu/g$, as well as the geometric mean of the local trap frequencies $\bar{\omega}$, using $\mu= \hbar \bar{\omega} \left(15 N a/\bar{a} \right)^{2/5}$, where $\bar{a} =\sqrt{\hbar/(m \bar{\omega})}$, $N=250$ is the atom number in the condensate, $a=5$~nm the scattering length, and $g=4\pi \hbar^2 a/m$. We obtain $n_{TF}=1.4(4) \times 10^{16}$cm$^{-3}$ for the peak condensate density and $\bar{\omega}/(2\pi)=21(6)$~kHz for the local trap frequency.

To verify the three-body loss hypothesis, we use the measured condensate release energy per particle $\varepsilon_r/h = 28$~kHz to calculate the chemical potential $\mu=(7/2) \varepsilon_r$ and atomic density \cite{rev_dalfovo}. Under the Thomas-Fermi approximation, we have for the peak density $n_{TF}=\mu/g$, where $g=4\pi \hbar^2 a/m$, with $a=100 a_0$ being the $^{87}$Rb scattering length, $a_0$ the Bohr radius, and $m$ the atomic mass. We find $n_{TF}=1.4 \times 10^{16}$~cm$^{-3}$, corresponding to a three-body loss rate $\Gamma_3 = K_3 \ n_\textrm{TF}^2 = 8 (3) \times 10^3~$s$^{-1}$. Here, we have used the loss coefficient 
$K_3 = 4(2) \times 10^{-29}~$cm$^{6}$s$^{-1}$ for distinguishable atoms, assuming that condensates are formed in all three magnetic sublevels of the $F=1$ manifold. The calculated loss rate is consistent with the observed loss rate of $\Gamma_{\textrm{obs}} = 4.3(3) \times 10^3~$s$^{-1}$ for the condensate during TOF. However, as discussed in the main text, the persistence of the condensate and thermal cloud in-trap for hundreds of milliseconds suggests we are overestimating the density in the condensate, and the most likely explanation for the loss of condensate atoms during TOF, as shown in Fig.~\ref{fig:ratio_volumes}c, are elastic two-body collisions.

\subsection{Atom number fitting procedure}

To obtain the atom number from our absorption images, we restrict (by eye) the vertical (or horizontal) axis to an interval containing the high-density peak(s) within the cigar-shaped cloud. By integrating along the other axis, the bimodal distribution is clearly visible in the resulting one-dimensional profile. For images that appear to contain only one BEC, we fit the resulting data to the sum of two Gaussian distributions, which describes a high-density component on top of a broad thermal component. The atom numbers within each component are then determined using the fit parameters (specifically, the optical depths and the RMS sizes). The residuals between the raw data and the bimodal fits give an estimate for the errorbars on the estimated atom number in the BEC.

For images that appear to contain multiple BECs, we fit the broad thermal component to a single Gaussian distribution, which is then subtracted from the data. The residuals depicting the high-density peaks are assumed to be BEC atoms, whose spatial extent is distinguished (by eye) from the remaining density fluctuations of the broad thermal background. These residuals are then integrated to give the number of atoms in the BEC. The main source of systematic errors is dominated by the finite offset of these residuals, which depends on the Gaussian fit to the thermal cloud. We assume the noise in this choice of offset is given by the RMS noise (standard deviation) on the background thermal distribution. 

\end{document}